\documentclass[acmtog]{acmart}
\acmSubmissionID{2424}

\usepackage{dblfloatfix}
\usepackage{multirow}
\usepackage{siunitx}
\usepackage[ruled]{algorithm2e}

\copyrightyear{2025} 
\acmYear{2025} 
\setcopyright{cc}
\setcctype{by}
\acmConference[arXiv Preprint Version]{arXiv Preprint Version}{October 11th}{2025}
\acmDOI{}
\acmISBN{}

\begin{document}
\title{TC-GS: A Faster Gaussian Splatting Module Utilizing Tensor Cores}

\author{Zimu Liao}
\orcid{0009-0000-4603-3914}
\affiliation{%
 \institution{Shanghai Jiao Tong University}
  \city{Shanghai}
 \country{China}}
\affiliation{%
 \institution{Shanghai Artificial Intelligence Laboratory}
  \city{Shanghai}
 \country{China}}
\author{Jifeng Ding}
\orcid{0009-0005-0975-4487}
\affiliation{%
 \institution{Shanghai Artificial Intelligence Laboratory}
  \city{Shanghai}
 \country{China}}

\affiliation{%
 \institution{Fudan University}
 \city{Shanghai}
 \country{China}}
\author{Siwei Cui}
\orcid{0009-0004-3544-723X}
\affiliation{%
 \institution{Shanghai Artificial Intelligence Laboratory}
    \city{Shanghai}
 \country{China}}
\affiliation{%
 \institution{Fudan University}
 \city{Shanghai}
 \country{China}}

\author{Ruixuan Gong}
\orcid{0009-0009-1303-0982}
\affiliation{%
 \institution{Beijing Institude of Technology}
 \city{Beijing}
 \country{China}}

\author{Boni Hu}
\orcid{0009-0001-4984-164X}
\affiliation{%
 \institution{Shanghai Artificial Intelligence Laboratory}
    \city{Shanghai}
 \country{China}}
\affiliation{
 \institution{Northwestern Polytechnical University}
 \city{Xi-an}
 \country{China}
}

\author{Yi Wang}
\orcid{0009-0000-9116-7698}
\affiliation{%
 \institution{Shanghai Artificial Intelligence Laboratory}
    \city{Shanghai}
 \country{China}}

\author{Hengjie Li}
\orcid{0009-0004-1986-1799}
\affiliation{%
 \institution{Shanghai Artificial Intelligence Laboratory}
    \city{Shanghai}
 \country{China}}
\affiliation{%
 \institution{Shanghai Innovation Insititute}
    \city{Shanghai}
 \country{China}}
 \email{lihengjie@pjlab.org.cn}
\authornotemark[1]
 
\author{Hui Wang}
\orcid{0009-0002-0604-4621}
\affiliation{%
 \institution{Shanghai Artificial Intelligence Laboratory}
    \city{Shanghai}
 \country{China}}

\author{Xingcheng Zhang}
\orcid{0009-0006-8525-0608}
\affiliation{%
 \institution{Shanghai Artificial Intelligence Laboratory}
    \city{Shanghai}
 \country{China}}

\author{Rong Fu}
\orcid{0009-0001-4549-8119}
\affiliation{%
 \institution{Shanghai Artificial Intelligence Laboratory}
  \city{Shanghai}
 \country{China}}
\email{furong@pjlab.org.cn}
\authornote{Corresponding Authors}

\begin{abstract}
3D Gaussian Splatting (3DGS) renders pixels by rasterizing Gaussian primitives, where conditional alpha-blending dominates the computational cost in the rendering pipeline. This paper proposes TC-GS, an algorithm-independ-ent universal module that expands the applicability of Tensor Core (TCU) for 3DGS, leading to substantial speedups and seamless integration into existing 3DGS optimization frameworks. The key innovation lies in mapping alpha computation to matrix multiplication, fully utilizing otherwise idle TCUs in existing 3DGS implementations. TC-GS provides plug-and-play acceleration for existing top-tier acceleration algorithms and integrates seamlessly with rendering pipeline designs, such as Gaussian compression and redundancy elimination algorithms. Additionally, we introduce a global-to-local coordinate transformation to mitigate rounding errors from quadratic terms of pixel coordinates caused by Tensor Core half-precision computation. Extensive experiments demonstrate that our method maintains rendering quality while providing an additional 2.18× speedup over existing Gaussian acceleration algorithms, thereby achieving a total acceleration of up to 5.6×.

\end{abstract}

\ccsdesc[500]{Computing methodologies~Rasterization}
\ccsdesc[300]{Computing methodologies~Parallel algorithms}

\keywords{Gaussian Splatting, Tensor Cores, Real-Time Rendering}

\begin{teaserfigure}
  \includegraphics[width=\textwidth]{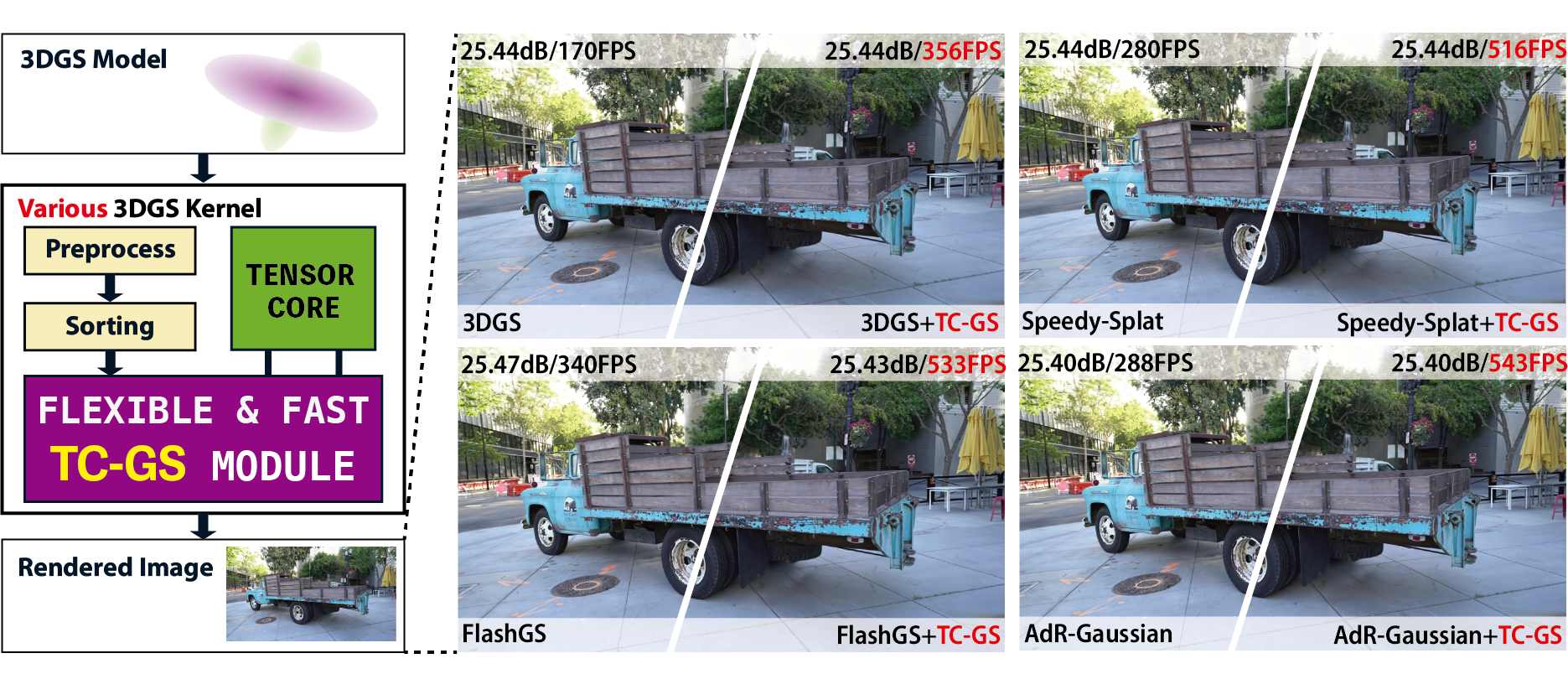}
  \caption{TC-GS: a Tensor Core-based acceleration module for 3D Gaussian Splatting that can be effortlessly applied to existing 3DGS rendering pipelines. Integrated into 3DGS and its variants, it achieves an extra $2\times$ speedup while preserving rendering quality.}
  \Description{Illustration of the TC-GS acceleration module applied to 3D Gaussian Splatting, showing speedup and rendering quality.}
  \label{fig:teaser}
\end{teaserfigure}

\maketitle

\section{Introduction}
\label{sec:introduction}

3D Gaussian Splatting (3DGS) represents a significant advancement in neural rendering, employing Gaussian primitives to achieve NeRF-comparable quality while reducing optimization time to tens of minutes~\cite{mallick2024taming,durvasula2023distwar}. However, 3DGS rendering speed is heavily constrained by extensive model parameters and inefficient rendering pipelines\cite{lu2024scaffold,ren2024octree,jiang2024vr}. 
Further acceleration remains essential for deployment on resource-constrained devices, real-time large-scale scene rendering systems, and low-latency edge computing scenarios. 

Despite numerous efforts to enhance computational efficiency \cite{wang2024adr, feng2024flashgs, hanson2024speedy, huang2025seele}, they fail to tap into the most powerful computing units in modern GPUs. Moreover, existing acceleration modules are tightly coupled with core components, preventing modular reuse and inter-method adaptation. This coupling further limits the practical deployment and scalability of 3DGS-based solutions.

Recent research \cite{wang2024adr, feng2024flashgs, hanson2024speedy, huang2025seele} has identified the \textit{conditional alpha-blending} process as the primary pipeline bottleneck. This process, which composites pixel colors with depth-sorted splatted Gaussian primitives, comprises three steps: (1) \textit{Alpha computation}: computing the alpha value (opacity) of each splatted Gaussian fragment on the pixel; (2) \textit{Culling}: eliminating fragments with low alpha values; and (3) \textit{Blending}: compositing remaining fragments into final pixel colors.

Many approaches~\cite{ye2025gsplat,wang2024adr,fan2024instantsplat} employ early culling strategies to reduce the range of splatted Gaussians, thus minimizing redundant alpha-blending computations. Although they eliminate redundant computation, these methods require corresponding pipeline designs to mitigate load imbalance issues introduced by precise intersection \cite{feng2024flashgs} calculations or sacrifice certain precision\cite{wang2024adr}. Additionally, we observe that over 80\% of Gaussian-pixel pairs that require culling remain in the pipeline. Consequently, the first two steps are the dominant part of the alpha-blending process.

Unlike previous efforts to remove redundant computing~\cite{wang2024adr, feng2024flashgs, hanson2024speedy, huang2025seele}, we focus on utilizing Tensor Core Units (TCUs) to accelerate alpha computation, which remains an unexplored area. In modern GPUs, TCUs are deployed to perform matrix multiply accumulate (MMA), i.e., $D=A\times B+C$ operations within a clock cycle, achieving high throughput on general matrix multiplication (GEMM) based computations including MLPs, CNNs, and transformers \cite{vaswani2017attention}. However, due to their specialized computation pattern, TCUs struggle to be applied to non-GEMM computations, leaving them idle in 3DGS.

To tackle this problem, this paper proposes TC-GS, a hardware-aligned redesign of alpha computation that unlocks Tensor Core acceleration for all 3DGS algorithms. TC-GS formulates a mapping from pixels and Gaussians into two matrices, where their matrix product yields logarithmic alpha values in batches. Implemented as a standalone module, TC-GS not only significantly improves rendering efficiency, but also enables seamless integration with existing 3DGS acceleration kernels, further enhancing overall performance.

In summary, the paper proposes the following contributions:

\begin{itemize}

\item A Detailed runtime analysis of the rendering pipeline showing that alpha computation in conditional alpha-blending is the performance bottleneck in 3DGS inference rendering.
\item The first use of Tensor Cores to accelerate alpha-blending, expanding the Tensor Core applicability beyond GEMM-based computations.
\item A global-to-local coordinate transformation (\textit{G2L}, Section \ref{sec:G2L}) that reduces the absolute magnitudes of quadratic coordinate terms, which are sensitive to Tensor Core half-precision computation, enabling lossless rendering quality while maintaining full acceleration benefits.
\end{itemize}

\section{Related Works}
\label{sec:related}

\subsection{Fast 3D Gaussian Splatting}
\label{subsec:related/3dgs}
3DGS \cite{kerbl20233d} has emerged as a leading real-time rendering technique with photo-realistic visual quality, finding wide applications diverse domains including autonomous driving \cite{zhou2024drivinggaussian}, robotics \cite{zhu20243dgsforrobtics}, and virtual reality \cite{jiang2024vr}. Current acceleration research explores two primary optimization strategies to enhance 3DGS performance. 

The first strategy modifies the algorithm itself to achieve better efficiency \cite{ye2025gsplat, mallick2024taming, lee2024compact}. CompactGS \cite{lee2024compact}, Mini-splatting \cite{fang2024mini}, PUP 3D-GS \cite{hanson2025pup}, and gsplat \cite{ye2025gsplat} introduce more efficient radiance field representations and adaptive resolution primitive pruning algorithms, while Taming-3DGS \cite{mallick2024taming} presents mathematically equivalent but computationally efficient solutions for gradient computation and attribute updates, substantially accelerating training speed. Stochastic rasterization for sorting-free 3DGS \cite{kheradmand2025stochasticsplats} is an another approach of improving the algorithm itself.

The second category focuses on improving computational efficiency of the original algorithm through intelligent task scheduling and CUDA kernel optimization \cite{feng2024flashgs, hanson2024speedy, wang2024adr, ye2025gsplat, gui2024balanced, durvasula2023distwar}. Balanced-3DGS \cite{gui2024balanced} optimizes the forward computation of rendering CUDA kernels within warps while minimizing thread idle time and maximizing resource utilization by evenly distributing tasks across computational blocks. Several works, including AdRGaussian \cite{wang2024adr}, FlashGS \cite{feng2024flashgs}, and SpeedySplat \cite{hanson2024speedy}, aim to reduce the number of pixels processed per Gaussian by designing more precise Gaussian-tile bounding boxes, with FlashGS \cite{feng2024flashgs} and SpeedySplat \cite{hanson2024speedy} further computing exact Gaussian-tile intersections. DISTWAR \cite{durvasula2023distwar} enhances processing efficiency through sophisticated thread management strategies, fully leveraging warp-level reductions in SM sub-cores and intelligently distributing atomic computations between SM and L2 atomic units based on memory access patterns. 
While these methods significantly accelerate training and rendering speed, tight coupling between algorithmic modules hinders transferability and reusability. Moreover, they fail to leverage modern hardware acceleration features such as Tensor Cores, further limiting the practical deployment and scalability of 3DGS solutions.

\subsection{Tensor Cores for Non-GEMM Based Computations}
\label{subsec:related/tc}
Tensor Cores Units are specialized hardware units developed by NVIDIA to accelerate Deep Learning applications~\cite{A100whitepaper}.
Each Tensor Core can perform a matrix multiply accumulate operation ${D = A \times B + C}$ on small matrices within a GPU cycle, where A and B must be in half percision format while the accumulators, C and D, can be either single or half percision. Tensor Core can achieve higher throughput than normal CUDA Cores: for instance on the NVIDIA A100 GPU, CUDA Cores has maximum 78 TFLOPS fp16 performance, whereas Tensor Cores can achieve 312 TFLOPS which is 4$\times$ times faster than CUDA Cores.

Despite this theoretical advantage, it is not a trival task to exploit Tensor Cores in arbitrary applications since Tensor Cores only support the matrix–multiply–accumulate (MMA) instruction rather than the full range of CUDA instructions set. This restriction requires algorithms to be reformulated in terms of small matrix operations, but the reward is exceptionally fast execution. Consequently, the strict requirements yet remarkable potential speed of Tensor Core operations have attracted researchers to explore adapting non-machine learning algorithms to GPU Tensor Cores, leveraging their performance potential. Previous works have successfully accelerated primitives such as reduction, scan and breadth-first search (BFS) using Tensor Cores \cite{dakkak2019accelerating,10.1145/3330345.3331057,niu2025berrybeesTC}. Recently, ConvStencil \cite{ConvStencil} has managed to adapt stencil computation to TCUs, while other works have explored TCU-based implementations in applications such as HighQR \cite{leng2024highQR} and TCUDB \cite{hu2022tcudb}. However, to our best knowledge, no prior work has explored how to exploit Tensor Cores directly within a 3D Gaussian Splatting rendering pipeline.


\begin{table}[t!]
  \centering
  \caption{Tensor Core \& CUDA Core Comparision among various GPUs. \textsuperscript{*} stands for using new sparse feature}
  \label{tab:fp16_perf}
  \begin{tabular}{lcc}
    \toprule
    \textbf{GPU} & \textbf{Peak FP16} & \textbf{Peak FP16 Tensor Core} \\
    \midrule
    H100 (PCIe) & 96 TFLOPS & 800 TFLOPS | 1600 TFLOPS\textsuperscript{*} \\
    A100        & 78 TFLOPS & 312 TFLOPS | 624 TFLOPS\textsuperscript{*} \\
    V100        & 31.4 TFLOPS & 125 TFLOPS \\
    RTX 4090    & 82.6 TFLOPS & 330.3 TFLOPS | 660.6 TFLOPS\textsuperscript{*} \\
    \bottomrule
  \end{tabular}
\end{table}

\section{Preliminaries and Observations}
\label{sec:preliminaries}

\subsection{3DGS Rendering Pipeline}
\label{subsec:preliminaries/3dgs}
In this paper, we focus on the official implementation of 3DGS, which is the mainstream pipeline widely used in most state-of-the-art acceleration methods.

A 3DGS model consists of a set of 3D Gaussians primitives $\mathcal{G}=\{G_1,G_2,\cdots,G_P\}$. Each Gaussian $G:\mathbb{R}^3\to\mathbb{R}$ is parameterized by its mean $\mu \in \mathbb{R}^3$, semi-definite covariance $\Sigma\in\mathbb{R}^{3\times 3}$, opacity $o\in(0,1]$, and color $c\in[0,1]^3$ and maps any 3D position $x\in\mathbb{R}^3$ to its density:
\begin{equation}
\label{equation:preliminaries/3dgs_distributions}
G(x)=o\mathrm{e}^{-\frac{1}{2}(x-\mu)^T\Sigma^{-1}(x-\mu)}. 
\end{equation}

The inference procedure of 3DGS can be divided into the following 3 stages:
\paragraph{Preprocess}
    The preprocess stage projects 3D Gaussian $G$ into 2D planes in parallel according to the viewing transform $W$ of a given camera:
\begin{equation}
    \mu^\prime=W\mu\in\mathbb{R}^2,\quad\Sigma^\prime=JW\Sigma W^TJ^T\in\mathbb{R}^{2\times2},
\end{equation}
    where $J=\frac{\partial{\mu'}}{\partial\mu}$ is the Jacobian of projection $W$. $\mu'=(\mu_x',\mu_y')$ and $\Sigma'$ is the mean and the covariance of projected Gaussian $G'$. The complexity of the preprocessing stage is $\mathcal{E}_p = O(P)$, where $P$ is the number of Gaussians.
    
\paragraph{Sorting}
    The screen space $\mathcal{S}$ are divided into disjoint tiles $\mathcal{T}$ with equal sizes. For each projected Gaussian, this stage identifies all the tiles it covers. Then, for each tile $\mathcal{T}\in\mathcal{S}$, the overlapped Gaussians are sorted by depth. A tile $\mathcal{T}$ coupled with a covering Gaussian $G'$ is denoted as a \textbf{splat} $(\mathcal{T},G')$. The total number of splats given by
\begin{equation}        N=\sum_{j=1}^P\sum_{\mathcal{T}\in\mathcal{S}}(\mathcal{T}\cap G'_j\neq\emptyset)=\sum_{\mathcal{T}\in\mathcal{S}}n(\mathcal{T}).
\end{equation}

where $\mathbf{1}$ is the indicator function that checks whether the Gaussian covers the tile. The process of determining coverage depends on the specific algorithm used. The complexity of the sorting stage is $\mathcal{E}_s = O(sN)$, where $s=64$ is the coefficient in radix-sort.


\begin{algorithm}
\caption{Conditional Alpha-blending}
\SetAlgoNoLine 
\KwIn{Tile $\mathcal{T}$, Sorted Projected Gaussians $\mathcal{G'(\mathcal{T})}=(G^{\prime}_1,\cdots,G^{\prime}_{n(\mathcal{T})})$}
\KwOut{Pixel colors $C_\mathcal{T}=(C_1,\cdots,C_{|\mathcal{T}|})$}
\ForEach{pixel $p \in \mathcal{T}$ \textbf{in parallel}}{
    $T \gets 1$ \tcp*{Initialize transmissivity} 
    $C \gets (0, 0, 0)$ \tcp*{Initialize RGB}
    \ForEach{Gaussian $G'\in\mathcal{G'(T)}$}{
        $\alpha \gets o\exp\left(-\frac{1}{2}(\mu^\prime-p)^T\Sigma^{\prime -1}(\mu^\prime-p)\right)$
        
        \If{$\alpha<\frac{1}{255}$}{
            \textbf{continue} \tcp*{Cull invisible Gaussians}
        }   
        \If{$T-\alpha{T}<0.0001$}{
            \textbf{break} \tcp*{Terminate blending}
        }
        $C\gets C + c\alpha T$ \tcp*{Composite colors}
        $T\gets T - \alpha T$ \tcp*{Update transmissivity}
    }
}
\label{algorithm:alpha_blending}
\end{algorithm}

\paragraph{Conditional Alpha-blending} This stage renders the final color of each pixel in a tile $\mathcal{T}=(p_1,p_2,\cdots,p_{|\mathcal{T}|})$ in parallel as Algorithm \ref{algorithm:alpha_blending}. Initially, for each projected overlapping Gaussian, the renderer will perform \textbf{alpha computation} to get the local opacity $\alpha$ at each pixel $p=(p_x,p_y)\in\mathbb{R}^2$:
\begin{equation}
    \alpha_{i,j}=o_j\exp{\left(-\frac{1}{2}(\mu^{\prime}_j-p_i)^T(\Sigma^{\prime}_j)^{-1}(\mu^{\prime}_j-p_i)\right)}.\label{equation: alpha_raw}
\end{equation}
Next, the renderer performs \textbf{culling} of Gaussians at each pixel if the local opacity is below a threshold:
\begin{equation}
    \alpha_{i,j}<\dfrac{1}{255}, \label{formula: vanilla pruning}
\end{equation}
The culling process is only a conditional judgment.

Otherwise, the surviving Gaussians will perform \textbf{blending} and contribute the pixel's color:
\begin{equation}
C_i=\sum_{j=1}^{n(\mathcal{T})}c_j\alpha_{i,j}T_{i,j};\quad T_{i,j}=\prod_{k=1}^{j-1}(1-\alpha_{i,k}). \label{formula: alpha-blending}
\end{equation}
where $T_{i,j}$ denotes the transmissivity and $n(\mathcal{T})$ is the number of splats of the tile $\mathcal{T}$. If $T_{i,j} <0.0001$, the renderer will \textbf{terminate} the blending process for that pixel and output the color. 
A pixel $p$ rendered by a Gaussian $G$ is referred to as a \textbf{fragment}, $(p, G)$. The complexity of the alpha-blending stage is $\mathcal{E}_a=O(|\mathcal{T}|N)$.

Thus, the total complexity of the Gaussian rasterization pipeline can be summarized as
$\mathcal{E}=\mathcal{E}_p+\mathcal{E}_s+\mathcal{E}_a$,
where \(\mathcal{E}_p, \mathcal{E}_s, \mathcal{E}_a\) represent the complexities of preprocess, sorting, and alpha-blending stages, respectively.

\subsection{Observation and Motivation}
\label{subsec:preliminaries/runtime}
\begin{figure*}[h]
    \centering
    \includegraphics[width=\linewidth]{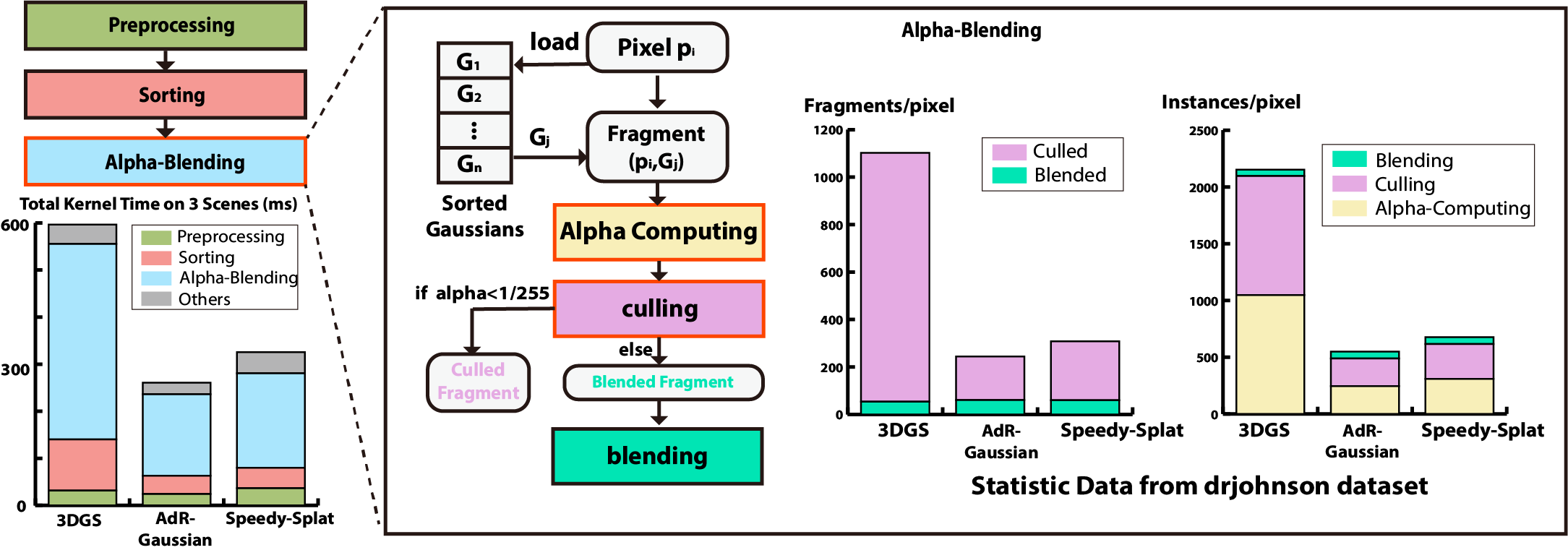}
    \caption{Analysis of 3DGS rendering bottlenecks: The left side shows the time distribution of preprocessing, sorting, and alpha-blending for 3DGS, AdR-Gaussian, and Speedy-Splat, with alpha-blending dominating. The right side details alpha computation, culling, and blending, identifying culled fragments as the primary bottleneck due to redundant alpha computations, while skipped fragments incur no cost. }
    \label{fig:overal}
\end{figure*}

To better understand the bottlenecks in current rendering methods, we conduct a detailed time breakdown analysis of the three stages mentioned above. Specifically, we profile 3DGS, AdR-Gaussian, and Speedy-Splat on an NVIDIA A800 using multiple datasets. While recent methods incorporate early culling strategies with more precise intersection checks to reduce the number of splats $N$, thereby decreasing the time spent on sorting and blending, the alpha blending stage still remains the dominant contributor to the overall rendering time, as shown in the bottom-left corner of Figure \ref{fig:overal}.


Motivated by this observation, we further analyze the core kernel \verb|renderCUDA|, which implements the alpha blending stage, in order to uncover deeper inefficiencies and guide targeted improvements.


\begin{figure}[h]
    \centering
    \includegraphics[width=0.99\linewidth]{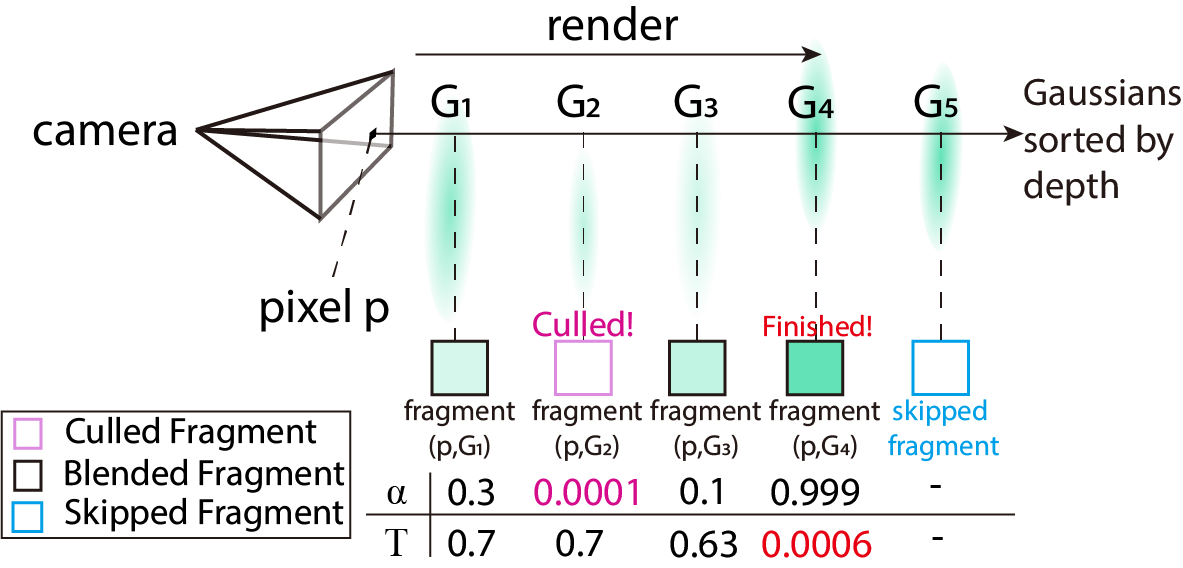}
    \caption{a) Three types of fragments on a pixel. b) If the Gaussian only covers a small portion of tile, a large amount culled fragments are generated.}
    \label{fig:fragmenttype}
\end{figure}

The alpha-blending stage will blend all fragments into pixel colors. These fragments are categorized into 3 types as Figure \ref{fig:fragmenttype} shows:
\begin{itemize}
    \item Culled fragments: Fragments that are discarded because $\alpha<\frac{1}{255}$. As shown in Figure \ref{fig:fragmenttype}, the second fragment is culled.
    \item Blended fragments: Fragments that contribute to the final pixel colors. As shown in Figure \ref{fig:fragmenttype}, the first, third, and fourth fragments are blended.
    \item Skipped fragments: Fragments that are ignored when $T<0.0001$, causing the renderer to terminate the blending process for the corresponding pixel. As shown in Figure \ref{fig:fragmenttype}, the fifth fragment and the ones that follow are skipped. There is \textbf{no} computation on skipped fragments.
\end{itemize}

\begin{table}[h]
    \centering
    \caption{The alpha computation and culling steps are run on both blended and culled fragments, while the blending step is only performed on blended fragments. There is no computation on skipped fragments.}
    \begin{tabular}{c|ccc}
         Fragments & Alpha Computation & Culling & Blending \\
         \hline
         Blended& \checkmark & \checkmark & \checkmark \\
         Culled&\checkmark & \checkmark & \\
        Skipped& & & \\
    \end{tabular}
    \label{tab:fragments category}
\end{table}

To model the computation for each type of fragments, we denote $k_\alpha, k_{\text{cull}}, k_{\text{blend}}$ as the average computation cost for alpha computation, culling, and blending. Similarly, let $f_{\text{blend}}, f_{\text{cull}}, f_{\text{skip}}$ represent the number of blended, culled and skipped fragments. According to Table \ref{tab:fragments category}, the total computation amount is:
\begin{equation}
\mathcal{C}=k_{\text{blend}}f_{\text{blend}}+(k_\text{cull}+k_\alpha)(f_\text{blend}+f_\text{cull}). \label{equation: performance model}
\end{equation}

To identify the bottleneck within the alpha-blending stage, we collect and categorize the generated fragments on various scenes using 3DGS, AdR-Gaussian and Speedy-Splat. As shown on the right side of Figure \ref{fig:overal}, culled and skipped fragments account for a large proportion. Since skipped fragments do not incur any computational cost, we can preliminarily conclude that the bottleneck lies in the alpha computation and culling steps.

\section{Method}
\label{sec:method}
\begin{figure*}[htbp]
    \centering
    \includegraphics[width=0.99\linewidth]{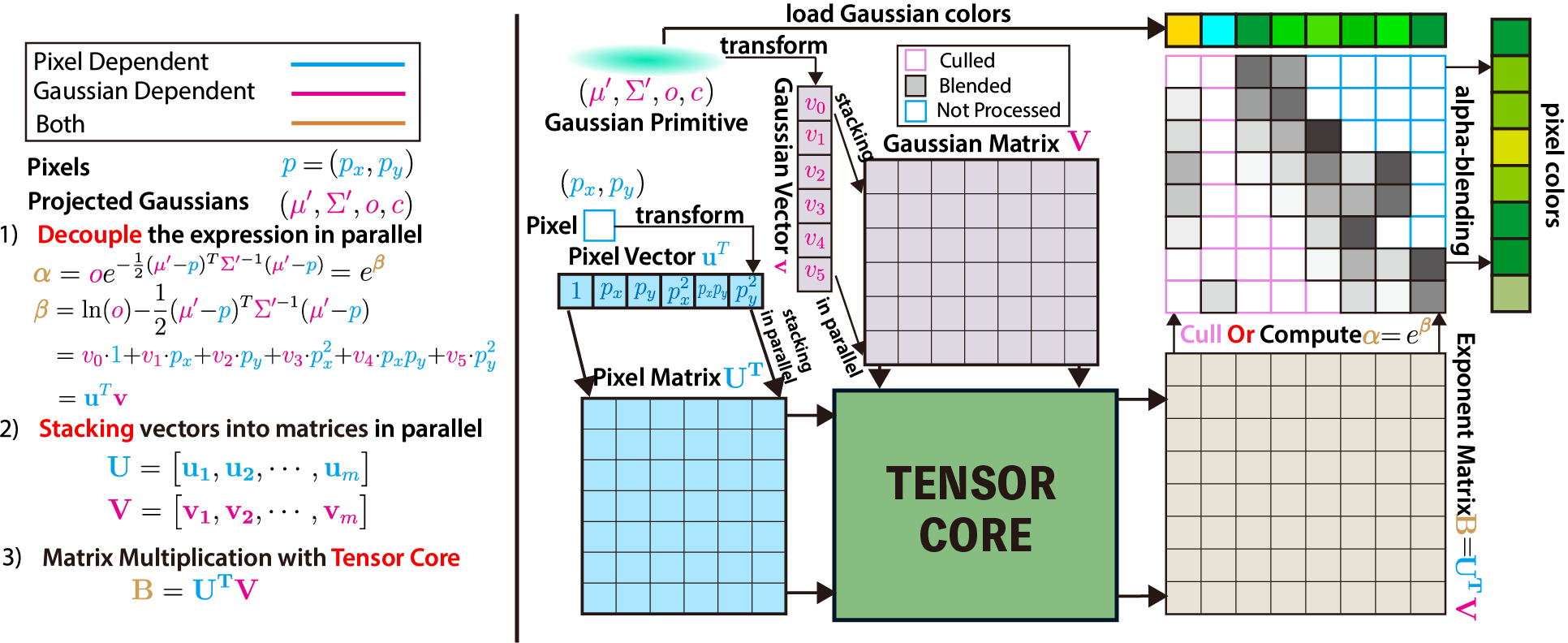}
    \caption{Design of \textit{Frag2Mat}: the alpha computation is reformulated as matrix multiplication to fully leverage Tensor Cores for accelerating alpha calculation.}
    \label{fig:design}
\end{figure*}

This section introduces how TC-GS utilizes the Tensor Cores to accelerate the alpha-computation and culling process, which are the bottlenecks of rendering. 
The fundamental components of TC-GS include \textit{EarlyCull} for culling optimization, \textit{Frag2Mat} for matrix transformation and \textit{G2L} for coordinate transformation. The detailed design is shown in Figure \ref{fig:design}.

\subsection{\textit{EarlyCull}: Reducing Exponential Computation}
First, TC-GS adopts \textit{EarlyCull}, an early-pruning strategy: it identifies invisible Gaussians that can be culled before the alpha-computation stage, greatly reducing the number of exponential instructions {\texttt{exp}} executed and improving throughput.

We begin by expressing Equation (\ref{equation: alpha_raw}) as the following form:
\begin{equation}
    \alpha_{i,j}=\mathrm{e}^{\beta_{i,j}} \label{equation: alpha_exp},
\end{equation}
where the exponent $\beta_{i,j}$ satisfies
\begin{equation}
    \beta_{i,j} =\ln(o_j)-\frac{1}{2}(\mu'_j-p_i)^T(\Sigma'_j)^{-1}(\mu'_j-p_i).
\label{equation: exponents}
\end{equation}
Then the culling condition, Equation (\ref{formula: vanilla pruning}), can be converted into 
\begin{equation}
    \beta_{i,j}<-\ln(255).
\end{equation}

\textit{EarlyCull} allows us to filter out non-contributing Gaussians in advance, reducing the amount of expensive {\texttt{exp}} instructions issued by alpha computation. Since prior experiments show that culled fragments account for a large portion of the workload, this approach yields a substantial speed-up for the entire pipeline.

\subsection{\textit{Frag2Mat}: Fitting the MMA Operation}
\label{subsec: frag2mat}

While Tensor Cores specialize in matrix multiply-accumulate (MMA) operations, the original alpha-computation involves per-pixel-per-Gaussian evaluations that exhibit non-coalesced memory access patterns and fragmented arithmetic operations – characteristics mismatched with matrix-oriented hardware architectures. To align with Tensor Cores' computational paradigm and fully exploit their high throughput, we restructure the alpha-computation into batched matrix operations through algebraic reformulation.

We propose \textit{Frag2Mat}, a computational restructuring that transforms per-pixel-per-Gaussian exponent calculations into unified matrix operations. By expressing the exponent as a linear combination of pixel coordinates, Gaussian parameters and their quadratic terms, we decouple multivariate dependencies into two independent components: a pixel vector encoding coordinate polynomials for its pixel, and a Gaussian vector composed of parameters from its Gaussian. The final exponent values are batch-calculated through standard matrix multiplication of two matrices formed by stacking pixel vectors and Gaussian vectors respectively.
LC
The core idea of TC-GS is decoupling the computation of exponents $\beta_{i,j}$ into two separated terms: one that depends only on the pixel index and another that depends only on the Gaussian. Notably, we observe that the exponent can be expressed as a quadratic function of the pixel $p$. Based on this observation, TC-GS transforms Equation (\ref{equation: exponents} into the following standard from:
\begin{equation}
\beta=v_{0} + v_1p_x + v_2p_y + v_3p_x^2+v_4p_{x}p_{y}+v_5p_{y}^2,\label{equation:expanded_dot_poduct}
\end{equation}
where
\begin{equation}
\begin{aligned}
v_0&=\ln(o)-\frac{1}{2}\mu^{\prime T}\Sigma^{\prime -1}\mu^{\prime},\\ v_1&=\sigma_{11}\mu'_x+\sigma_{12}\mu'_y,\quad v_2=\sigma_{12}\mu'_x+\sigma_{22}\mu'_y,\\
v_3&=-\dfrac{1}{2}\sigma_{11},\quad v_4=-\sigma_{12},\quad v_5=-\dfrac{1}{2}\sigma_{22}\\
\end{aligned}
\end{equation}
with the notation that symmetric $\Sigma'^{-1}=\begin{bmatrix}
    \sigma_{11}&\sigma_{12}\\\sigma_{21}&\sigma_{22}
\end{bmatrix}$ with $\sigma_{12}=\sigma_{21}$.

The Equation (\ref{equation:expanded_dot_poduct}) is the expanded dot product of two vectors:
\begin{equation}
    \beta_{i,j}=\mathbf{u}_i^T\mathbf{v}_j, \label{equation: dotprod}
\end{equation}
where
\begin{equation}
    \mathbf{u}=\begin{bmatrix}
    1&p_{x}&p_{y}&p_{x}^2&p_{x}p_{y}&p_{y}^2 
    \end{bmatrix}^T \label{equation:define vec u}
\end{equation}
and
\begin{equation}
    \mathbf{v}=\begin{bmatrix}
        v_0&v_1& v_2& v_3& v_4& v_5
    \end{bmatrix}^T.
\end{equation}
Here $\mathbf{u}$, the pixel vector solely dependent on pixel properties, which will be computed only \textit{once} for each pixel. Similarly, each Gaussian vector $\mathbf{v}$ can be only computed once. Therefore, the expression of the exponent is decoupled into the dot product of two individual vectors. 
To align the computation to the Tensor Cores, TC-GS stacks the vectors into two independent matrices:
\begin{equation}
    \mathbf{U}=\begin{bmatrix}
\mathbf{u}_1&\mathbf{u}_2&\cdots&\mathbf{u}_m
    \end{bmatrix}\in\mathbb{R}^{6\times m},
\end{equation}
\begin{equation}
    \mathbf{V}=\begin{bmatrix}
\mathbf{v}_1&\mathbf{v}_2&\cdots&\mathbf{v}_j
    \end{bmatrix}\in\mathbb{R}^{6\times n},
\end{equation}
where $m$ is the the number of pixels within a tile, and $n$ is the number of Gaussians whose projection is overlapped with the tile. Then TC-GS computes the exponent matrix $\mathbf{B}\in\mathbb{R}^{m\times n}$ as matrix multiplication:
\begin{equation}
    \mathbf{B}=\begin{bmatrix}
        \beta_{1,1}&\cdots&\beta_{1,n}\\
        \vdots&\ddots&\vdots\\
        \beta_{m,1}&\cdots&\beta_{m,n}\\
    \end{bmatrix}=
    \begin{bmatrix}
        \mathbf{u}_1^T\mathbf{v}_1&\cdots&\mathbf{u}_1^T\mathbf{v}_n\\
        \vdots&\ddots&\vdots\\
        \mathbf{u}_m^T\mathbf{v}_1&\cdots&\mathbf{u}_m^T\mathbf{v}_n\\
    \end{bmatrix}
    =\mathbf{U}^T\mathbf{V}.\label{equation:MM}
\end{equation}

Therefore, we can leverage Tensor Cores to compute $\beta$ for $m$ pixels and $n$ Gaussians simultaneously. As Tensor Cores natively support half-precision matrix operations, additional design is required to ensure numerical stability throughout the alpha computation.

\subsection{\textit{G2L}: Fitting the Half Precision}
\label{sec:G2L}

Although we adopted Tensor Cores for alpha computation and culling. The challenges remain as  Tensor Cores do not support common single-precision floating-point operations. Instead, they only take FP16 or TF32 as the input of the MMA operation, resulting in a higher machine epsilon of $\varepsilon_H=9.77\times 10^{-4}$ compared to FP32's $\varepsilon_F=1.19\times10^{-7}$. The rounding error of a floating number $r$ is $|r|\varepsilon_H$. Therefore, the error of computation on Tensor Cores is more sensitive at the absolute value of the input data.

We notice that the range of the pixel $p$ is $[0,w]\times[0,h]$, where $w, h$ are the width and height of the screen space, respectively. When loading the pixel vector $\mathbf{u}$ in Equation (\ref{equation:define vec u}) into Tensor Cores, its absolute values of quadratic terms $p_x^2$, $p_xp_y$ and $p_y^2$ can exceed a million, contributing catastrophic rounding errors to $\beta$. Moreover, overflow will occur if the FP16 range $[-6.55\times10^5, 6.55\times10^5]$ is used.

\begin{figure}[h]
    \centering
    \includegraphics[width=\linewidth]{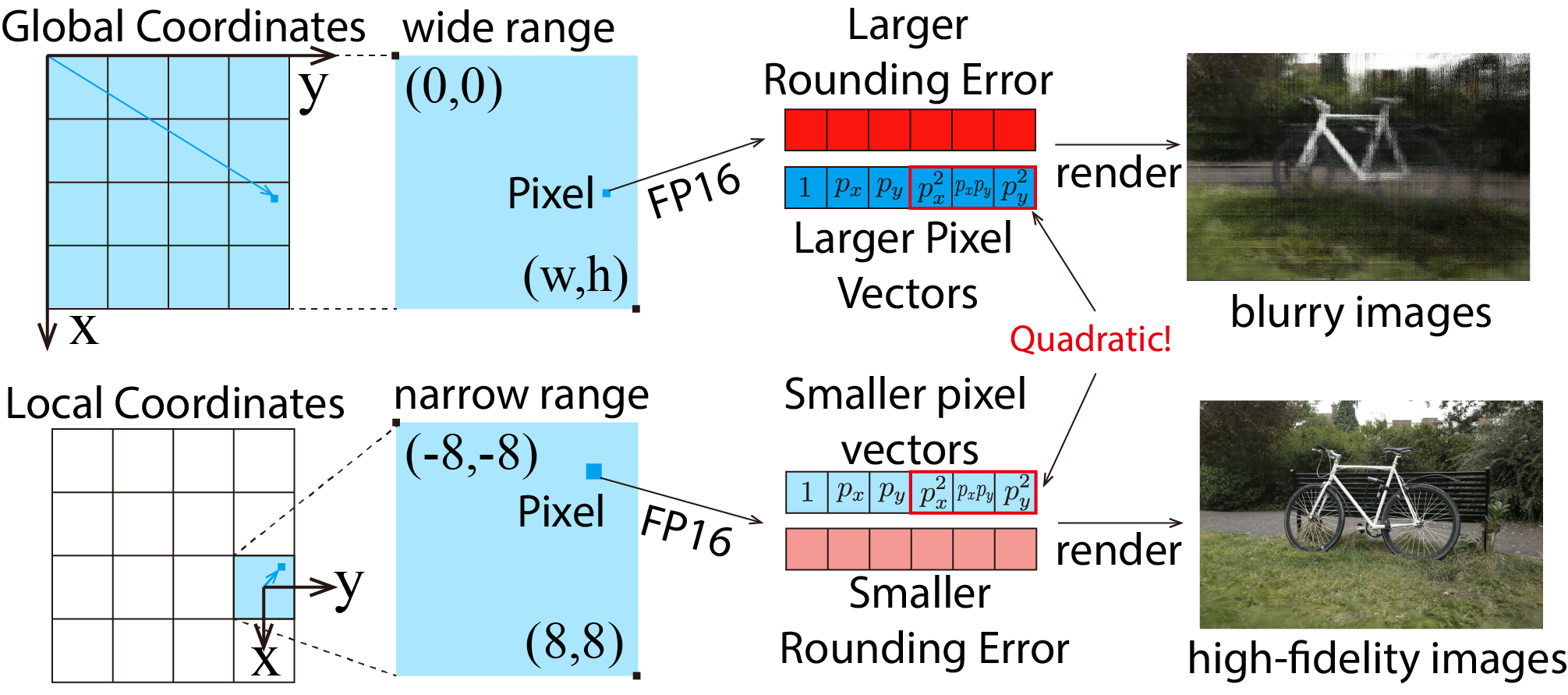}
    \caption{The quadratic term of pixel coordinates significantly contributes the rounding error, resulting in blurry images. Local coordinates constrain the value of $\Delta p$ into $[-8,8]^2$, which can reduce the upper bound of the rounding error.}
    \label{fig:local-coordinates}
\end{figure}

To tackle this problem, TC-GS has designed \textit{G2L}, a coordinate mapping, which significantly reduces the absolute value of terms in pixel vectors. 
Since all pixels $p$ within a tile $\mathcal{T}$ will be rendered in parallel, \textit{G2L} can map the coordinates of pixels into the local coordinate:
\begin{equation}
    \Delta p=p-p(\mathcal{T}),
\end{equation}

Since each tile contains $16\times16$ pixels, then the range of pixel in LC is $\Delta p\in[-8,8]^2$, and the maximum element of $\mathbf{u}_\Delta$ will not exceed $64$. This conversion does not only avoid the overflow in FP16, but also significantly reduces the rounding error.

Respectively, the means of Gaussians in the local coordinate is 
\begin{equation}
    \Delta \mu^\prime=\mu^\prime-p(\mathcal{T}).
\end{equation}
Notice that the relative position between $\mu'$ and $p$ stays invariant:
\begin{equation}
    \Delta \mu'-\Delta p=\mu'-p.
\end{equation}
Therefore, \textit{Frag2Mat} can be also performed on local coordinates.

As the proof in the appendices goes, \textit{G2L} reduces the quadratic rounding error $O(h^2+hw+w^2)\varepsilon_H$ into linear $O(h+w)\varepsilon_H$, and achieve better rendering quality in FP16 as Section \ref{subsec:exp/ablation} demonstrates.

\section{Experiment}
\label{subsec:exp}

\subsection{Experiment Setup}
\label{sec:exp/setup}

\begin{table*}
  \centering
  \caption{Comparison With/Without TC-GS Module Across Datasets.}
  \label{tab:comparison}
  \resizebox{\textwidth}{!}{
  \begin{tabular}{lcccccccccccc}
  \toprule
  \multirow{2}{*}{Method} & 
  \multicolumn{4}{c}{Tanks \& Temple} &
  \multicolumn{4}{c}{Deep-Blending} &
  \multicolumn{4}{c}{Mip-NeRF360}\\
  \cmidrule(lr){2-5} \cmidrule(lr){6-9} \cmidrule(lr){10-13} 
    & FPS$\uparrow$ & PSNR$\uparrow$ & SSIM$\uparrow$ & LPIPS$\downarrow$ & FPS$\uparrow$ & PSNR$\uparrow$ & SSIM$\uparrow$ & LPIPS$\downarrow$ & FPS$\uparrow$ & PSNR$\uparrow$ & SSIM$\uparrow$ & LPIPS$\downarrow$ \\
  \midrule
  3DGS & 161.548 & 23.687 & 0.851 & 0.169 & 131.373 & 29.803 & 0.907 & 0.238 & 128.88 & 26.546 & 0.785 & 0.25 \\
  3DGS(+TC-GS) & 323.423(2.13$\times$) & 23.682 & 0.851 & 0.169 & 286.819(2.185$\times$) & 29.803 & 0.906 & 0.236 & 258.809(2.01$\times$) & 26.544 & 0.785 & 0.25 \\
  FlashGS & 326.835 & 23.706 & 0.851 & 0.169 & 563.590 & 29.789 & 0.907 & 0.239 & 399.772 & 26.551 & 0.786 & 0.251 \\
  FlashGS(+TC-GS) & 539.387(1.65$\times$) & 23.684 & 0.851 & 0.169 & 735.777(1.305$\times$) & 29.806 & 0.906 & 0.236 & 479.338(1.199$\times$) & 26.508 & 0.785 & 0.25 \\
  Speedy-Splat & 268.602 & 23.687 & 0.852 & 0.17 & 315.822 & 29.803 & 0.907 & 0.238 & 264.23 & 26.546 & 0.785 & 0.25 \\
  Speedy-Splat(+TC-GS) & 521.284(1.94$\times$) & 23.684 & 0.852 & 0.169 & 579.988(1.84$\times$) & 29.802 & 0.906 & 0.236 & 465.308(1.76$\times$) & 26.544 & 0.785 & 0.25 \\
  AdR-Gaussian & 278.65 & 23.635 & 0.852 & 0.17 & 324.18 & 29.814 & 0.907 & 0.238 & 261.339 & 26.52 & 0.785 & 0.25 \\
  AdR-Gaussian(+TC-GS) & 545.154(1.955$\times$) & 23.632 & 0.851 & 0.169 & 612.063(1.89$\times$) & 29.815 & 0.906 & 0.236 & 467.310(1.78$\times$) & 26.519 & 0.785 & 0.25 \\
  \bottomrule
  \end{tabular}
  }
\end{table*}
\paragraph{Implementation} Current computation patterns which Tensor Cores support do not include length-6 MMA operations. A simple way to align the required length-8 MMA operation is zero padding. In our implementation, we replace $v_0$ in $\mathbf{v}$ with $\frac{v_0}{3}$ and duplicate it 3 times. The constant term $1$ in $\mathbf{u}$ are duplicated respectively. This approach can reduce the rounding error by reducing the absolute value.

\paragraph{datasets\&metrics}
To prove that our method is universal and efficient, we conduct experiments on the same datasets as those used in 3DGS\cite{kerbl20233d}. Specifically, the datasets contain all scenes from Mip-NeRF360\cite{barron2022mip}, two outdoor scenes (Truck \& Train) from Tanks \& Temple\cite{knapitsch2017tanks}, two indoor scenes (Drjohnson \& playroom) in Deep-Blending\cite{hedman2018deep}. For each scene, CUDA events are inserted at the start and end of the forward rendering procedure to measure FPS values, and the reported results represent the average metrics of the test datasets.All experiments are conducted on an NVIDIA A800 GPU (80GB SXM) equipped with 432 Tensor Cores, achieving a peak 624 TFLOPS FP16 performance.


\subsection{Comparisons}
\label{subsec:exp/comparisons}
TC-GS demonstrates significant performance improvements in all tested methods while keeping rendering quality. The qualitative and quantitative comparison between four baseline rendering pipelines and their results after integrating the TC-GS module are shown in Table \ref{tab:comparison}.

\paragraph{Rendering speed}
Across all datasets, incorporating the TC-GS module roughly \textbf{doubles} frame throughput except for FlashGS. The state-of-the-art performance is attained by FlashGS integrated with TC-GS, which achieves a \textbf{3.3–5.6$\times$} speedup compared to the original 3DGS, while delivering rendering speeds of \textbf{479.3–735.8} FPS.


By comparing the performance of alpha-blending, as shown in Figure \ref{renderCUDAComparison}, the rendering pipeline gains a 2-4$\times$ performance improvement. Combined with the previous analysis, this further emphasizes the importance of optimizing the rendering pipeline.

While AdR-Gaussian and Speedy-Splat already employ advanced optimizations to reduce redundant computation, TC-GS further pushed their performance boundaries, delivering significant speedup of \textbf{1.87$\times$} and \textbf{1.84$\times$}  respectively. As for FlashGS, while its existing pipeline rendering optimizations yield a more modest 1.38$\times$ average acceleration with TC-GS integration, we emphasize that the pipeline optimizations and TC-GS's approach operate orthogonally.

\paragraph{Image Quality}
Table \ref{tab:comparison} also demonstrates the rendered images quality metrics across all methods. We observe negligible differences in PSNR, SSIM, and LPIPS metrics between each method's original implementation and its TC-GS-enhanced counterpart. The minimal variations observed likely originate from hardware-level instruction set disparities and floating-point precision conversion artifacts in the rasterization pipeline.

\begin{table}[]
    \centering
    \caption{Comparison of alpha-blending time across methods and datasets.}
    \resizebox{\linewidth}{!}{
    \begin{tabular}{cccc}
    \toprule
    \textbf{Method} & \textbf{Scene} & \textbf{with TC-GS time(ms)} & \textbf{original time(ms)} \\
    \midrule
    \multirow{3}{*}{\textbf{3DGS}}     & drjohnson & 1.348(3.49$\times$)   & 4.705   \\
                  & train     & 1.217(3.57$\times$)    & 4.348       \\
                  & flowers   & 0.880(3.68$\times$)    & 3.236        \\
    \midrule
    \multirow{3}{*}{\textbf{FlashGS}} & drjohnson & 0.540(2.18$\times$)    & 1.179   \\
                  & train     & 0.609(2.03$\times$)    & 1.234       \\
                  & flowers   & 0.625(2.38$\times$)    & 1.487        \\
    \midrule
    \multirow{3}{*}{\textbf{AdR-Gaussian}} & drjohnson & 0.406(4.76$\times$)    & 1.931      \\
                  & train     & 0.518(4.23$\times$)    & 2.191             \\
                  & flowers   & 0.469(4.39$\times$)    & 2.061            \\
    \midrule
    \multirow{3}{*}{\textbf{Speedy-Splat}} & drjohnson & 0.510(3.86$\times$)    & 1.969    \\
                  & train     & 0.672(3.46$\times$)    & 2.326           \\
                  & flowers   & 0.592(3.52$\times$)     & 2.083           \\
    \bottomrule
    \end{tabular}
    \label{renderCUDAComparison}
    }
\end{table}

\subsection{Ablation Study}
\label{subsec:exp/ablation}


To determine the source of the acceleration, we conducted an ablation study on \textit{EarlyCull} and \textit{Frag2Mat} with \textit{G2L} across various renderers and datasets. As Table \ref{ablation:earlyCull} shows, both methods can accelerate the rendering speed. \textit{EarlyCull} alone achieves $1.06-1.51\times$ speedup, but it accelerates slightly when \textit{Frag2Mat} is enabled. \textit{Frag2Mat} alone achieves $1.97-2.24\times$ speedup, and it still  
accelerates significantly with \textit{EarlyCull} enabled. Therefore, \textit{Frag2Mat} with $\textit{G2L}$ contributes the main speedup.

\begin{table}[!ht]
\caption{Rendering FPS across settings, renderers, and datasets.
To evaluate the effect of each component in TC-GS, we conducted an ablation study
on EarlyCull and Frag2Mat with G2L.}
\label{tab:ablation_study}
\resizebox{\linewidth}{!}{
\begin{tabular}{@{}cccccc@{}}
\toprule
\multirow{2}{*}{\textit{EarlyCull}} & \multirow{2}{*}{\textit{Frag2Mat} with \textit{G2L}} &
\multicolumn{2}{c}{\textbf{3DGS}} & 
\multicolumn{2}{c}{\textbf{SpeedySplat}} \\
\cmidrule(lr){3-4} \cmidrule(lr){5-6}
 &  & truck & drjohnson & truck & drjohnson \\ 
\midrule
& & 169.834 & 108.123 & 366.457 & 349.673 \\ 
\checkmark &  & 210.110 & 163.967 & 406.434 & 371.605 \\ 
& \checkmark & 351.965 & 242.993 & 732.008 & 688.217 \\ 
\checkmark & \checkmark & 361.839 & 253.770 & 727.647 & 701.254 \\ 
\bottomrule
\end{tabular}
}
\label{ablation:earlyCull}
\end{table}

\begin{figure}[h]
    \centering
    \includegraphics[width=\linewidth]{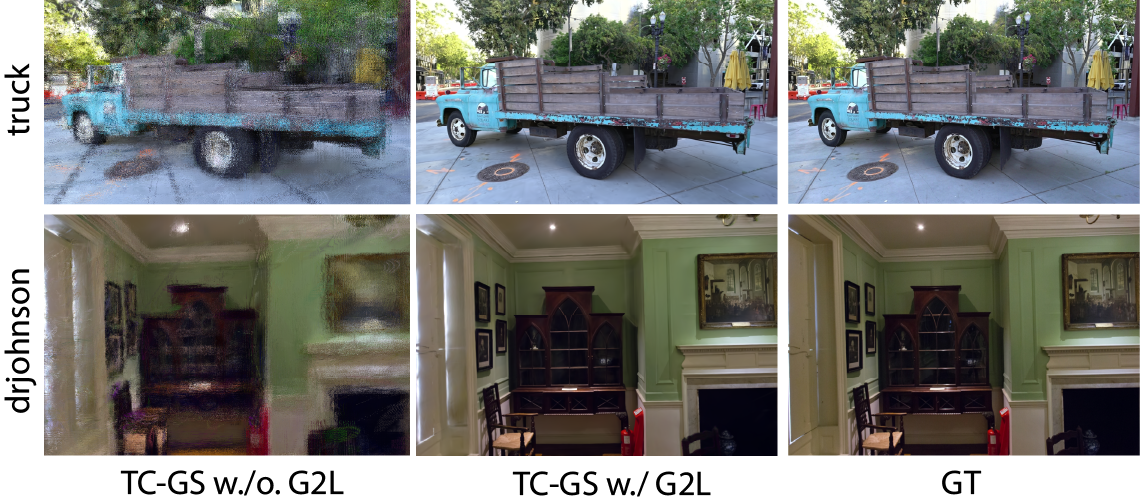}
    \caption{Ablation study on G2L when applying TC-GS on original 3DGS.}
    \label{fig:ablation}
\end{figure}

To clearly illustrate the impact of G2L on rendering quality and performance, we ran the following comparisons shown in Table \ref{ablation} and the rendering images are shown in Figure \ref{fig:ablation}. The ablation experiment shows that all Tensor Core variants boosted FPS. However, direct TF32 computation already introduced noticeable image-quality degradation, and FP16 alone dropped PSNR to just 8 dB. In contrast, combining FP16 Tensor Core optimization with Local Coordinates method preserved nearly same image quality while delivering the highest FPS. This demonstrates that our G2L can simultaneously maximize performance and maintain rendering fidelity.
\begin{table}[h]
    \centering
    \caption{Ablation study on applying TC-GS on original 3DGS.}
        \resizebox{\linewidth}{!}{
    \begin{tabular}{ccccccc}
    \hline
    \multirow{2}{*}{\textit{Frag2Mat}}&\multirow{2}{*}{\textit{G2L}} &\multirow{2}{*}{Precision}&\multicolumn{2}{c}{drjohnson}  & \multicolumn{2}{c}{truck} \\
    \cmidrule(lr){4-5} \cmidrule(lr){6-7} 
    &&&PSNR$\uparrow$&FPS$\uparrow$&PSNR$\uparrow$&FPS$\uparrow$\\
    \hline
    &&FP32& 29.48& 108.123 &25.44&169.834\\
    $\checkmark$&&FP16&8.85 & 253.79 & 6.24& 356.619\\
    $\checkmark$&&TF32&20.02& 181.429&14.27 &290.047 \\
    $\checkmark$&$\checkmark$&FP16&29.46& 253.77 &25.44&361.839\\
    \hline
    \end{tabular}
    }
    \label{ablation}
\end{table}

\subsection{Discussion}
\label{subsec:exp/discussion}
\subsubsection{Limitations}
While our method achieves significant acceleration in rendering computations, further optimization could be attained by adopting pipeline techniques similar to those proposed in FlashGS. However, unlike FlashGS's thread-level pipelining, a warp-level pipelining approach is required. Notably, post-optimization profiling reveals an increased proportion of preprocess overhead in the forward procedure. This preprocess phase inherently involves numerous matrix-based operations such as view transformations, providing opportunities for optimization using Tensor Cores.

\subsubsection{Training}
Although TC-GS is primarily designed to optimize real-time rendering, our implementation serves as a plug-and-play accelerator for both inference and training, requiring no modifications to the original 3DGS model or the training process.

\section{Conclusion}
\label{sec:conclusion}
In this paper, we propose TC-GS, a high-performance rendering module for accelerating 3D Gaussian Splatting. TC-GS optimizes the rendering pipeline by decomposing alpha-blending into two components, i.e., Gaussian-related and pixel-related matrices.
Through reformulating alpha-blending as matrix multiplications, TC-GS \newline achieves a significant speedup of 2.03$\times$ to 4.76$\times$ for alpha blending on Tensor Cores. When integrated into state-of-the-art pipelines, TC-GS attains approximately a 2$\times$ throughput improvement; even when compared against highly optimized, software-pipelined methods, it delivers a 1.38$\times$ speed-up. These gains are realized without any degradation in image quality, demonstrating TC-GS’s superior performance, scalability, and broad applicability for future real-time neural rendering systems.

\begin{acks}
This work is supported by the National Key R$\&$D Program of China (No. 2022ZD0160201), 
Shanghai Artificial Intelligence Laboratory, and Shanghai Municipal Science and Technology Major Project.
\end{acks}

\bibliographystyle{ACM-Reference-Format}
\bibliography{bibliography}

    `\newpage
\appendix
\section{Derivation and Proof of the Upper Bound of Rounding Error on TC-GS}

\label{proof}
\subsection{Floating-Point Numbers}
\label{subsec:fp}

In computing, floating-point numbers arithmetic is arithmetic on subset of real numbers, which can be represented with a finite amount of memory. A floating-point number $r$ is formed by a significand $s$, base $b$ and exponent $e$:
\begin{equation}
    r = s\cdot b^e.
\end{equation}

In this paper, we only consider the binary formats, whose base $b$ is $2$.

In most modern computing systems, a floating-point number is consists of 3 parts:
\begin{itemize}
    \item Sign bit: Represents the sign of the number. A sign bit of $0$ indicates a positive number, and $1$ indicates a negative number.
    \item Exponent: Represents the scale of the number. The exponent is stored with an offset. 
    \item Significand: Represents the mantissa of the floating-point number. The number of bits of significand is the precision.
\end{itemize}

For example, if the significand has $m$ bits, the real value of the floating-point number is:
\begin{equation}
    (-1)^{\text{sign}}(1.b_{m-1}b_{m-2}\dots b_0)_2\times 2^{\text{exponent - offset}},
\end{equation}

where $b_{m-1}b_{m-2}\dots b_0$ is the significand bits. Table \ref{tab:floating points} shows 3 floating-point formats in this paper. FP32 is also known as single-precision floating-point format, and FP16 is called half-precision floating-point format.
\begin{table}[H]
    \centering
    \begin{tabular}{c|cccc}
        \hline
         Format & Sign & Exponents & Significand & Machine Epsilon\\
         \hline
         FP16& $1$bit& $5$bits& $10$bits & $\varepsilon_H=9.77\times 10^{-4}$\\
         TF32& $1$bit& $8$bits& $10$bits & $\varepsilon_H=9.77\times 10^{-4}$\\
         FP32& $1$bit& $8$bits& $23$bits & $\varepsilon_F=1.19\times 10^{-7}$\\
         \hline
    \end{tabular}
    \caption{3 binary floating-point formats in this paper.}
    \label{tab:floating points}
\end{table}

\subsection{Machine Epsilon}
\textbf{Rounding} is a procedure for representation of a real number $r$ $(r\neq 0)$ into a floating-point format as $R(r)$, where $R$ is the rounding function. The \textbf{relative error} $\varepsilon$ of the rounding procedure is defined as:
\begin{equation}
    \varepsilon(r) = \dfrac{|R(r)-r|}{|r|}.
\end{equation}

For a fixed floating-point system and a rounding procedure, \textbf{machine epsilon} $\varepsilon_\text{mach}$ is an upper bound of relative error if no underflow or overflow occurs:
\begin{equation}
|R(r)-r|\leq\varepsilon_\text{mach}|r|. \label{equation: define mach epsilon}
\end{equation}

\proposition{If $\varepsilon_\text{mach}$ is a machine epsilon of a floating-point format, the rounded value of a real number $r$ satisfies:
\begin{equation}
    R(r)=r+O(\varepsilon_\text{mach}). 
\end{equation} \label{proposition 0}
}
\proof{
Since the $\varepsilon_\text{mach}$ does not depend on $r$, then Inequality (\ref{equation: define mach epsilon}) shows $|R(r)-r|=O(\varepsilon_\text{mach})$. Therefore, $R(r)-r=O(\varepsilon_\text{mach})$.
}

\proposition{If a binary floating-point format has $m$ significand bits, $\varepsilon_\text{mach}=2^{-m}$ is its machine epsilon.}\label{proposition 1}

\proof{
Considering $|r|\in[2^e,2^{e+1})$ is a real number which does not overflow or underflow in this format, it will lies between two adjacent floating-point numbers, $r_b$ and $r_u$:
\begin{equation}
    r_b\leq|r|< r_u.
\end{equation}

Suppose $x_b=(1.b_{m-1}b_{m-2}\dots b_0)_2\times 2^e$, then the adjacent $r_u$ is:
\begin{equation}
    r_u= (1.b_{m-1}b_{m-2}\dots b_0)_2 + (0.00\cdots1)_2\times 2^e = r_b+2^{e-m}.
\end{equation}

Since $R(r)$ is either $r_u$ or $r_b$,
\begin{equation}
    |R(r)-r|\leq |r_u-r_b|=2^{e-m}=2^e\cdot2^{-m}\leq 2^{-m}|r|.
\end{equation}

Therefore, $2^{-m}$ is a machine epsilon of the floating-point format.
}

From Proposition \ref{proposition 1}, we can conclude that the machine epsilon is only depend on the number of significand bits. 

In this paper, we represent the machine epsilon of TF32 and FP16 as $\varepsilon_H=9.77\times 10^{-4}$, and the machine epsilon of FP32 as $\varepsilon_F=1.19\times 10^{-7}$. Meanwhile, the rounding process of FP16 and TF32 is denoted as $R_H$, while the rounding process of FP32 is represented as $R_F$.

Furthermore, if the format is fixed, the rounding error is only depend on the absolute value $|r|$. The larger absolute value contribute the larger rounding error. The core idea of \textit{G2L} in Section 4.3 is reducing the absolute value during the computing process.



\subsection{Rounding Error of Dot Product with Tensor Cores}

Matrix multiplication can be interpreted as the dot product of vectors at each element, as discussed in Section 4.2. Therefore, to analyze the rounding error in matrix multiplication, it suffices to examine the error in the dot product:
\begin{equation}
    \beta=\mathbf{u}^T\mathbf{v}=\sum_{i=1}^nu_iv_i,
\end{equation}
where $n$ is the length of the vectors.

The dot product can be decomposed as a multiplication step and a accumulation step. On Tensor Cores, the multiplication step are performed on half-precision format (FP16 or TF32) and the accumulation step can be executed on FP32 format. 

We omit the detailed analysis on terms of small $\varepsilon_H^2$ and $\varepsilon_F$ and represent them as $O(\varepsilon_H^2)$ and $O(\varepsilon_F)$.

\proposition{On Tensor Cores, the absolute error of $\beta=\mathbf{u}^T\mathbf{v}$ satisfies
\begin{equation}
    |\hat\beta-\beta|\leq2\varepsilon_H\sum_{i=1}^n|u_iv_i|+O(\varepsilon_H^2+\varepsilon_F),
\end{equation}where $\hat\beta$ is the computed value of $\beta$ with Tensor Cores. \label{proposition: dot product}
}

\proof{
At first, we consider product of $u_i$ and $v_i$. Before multiplication, these two real number are stored in half-precision format as $R_H(u_i)$ and $R_H(v_i)$ satisfying
\begin{equation}
    |R_H(u_i)|\leq|R_H(u_i)-u_i| + |u_i|\leq|u_i|\varepsilon_H+|u_i|=(1+\varepsilon_H)|u_i|
\end{equation}
and
\begin{equation}
    |R_H(v_i)|\leq(1+\varepsilon_H)|v_i|.
\end{equation}

Therefore,
\begin{equation}
    |R_H(u_i)R_H(v_i)|\leq(1+\varepsilon_H)^2(|v_i||u_i|)=(1+2\varepsilon_H)|u_iv_i|+O(\epsilon_H^2).
\end{equation}

The value product are stored in FP32 format for next accumulation step on Tensor Cores:

\begin{equation}
\begin{aligned}
    |R_F(R_H(u_i)R_H(v_i))|&\leq(1+\varepsilon_F)|R_H(u_i)R_H(v_i)| \\
    &\leq(1+\varepsilon_F)((1+2\varepsilon_H) |u_iv_i|+O(\varepsilon_H^2)) \\
    &\leq (1+2\varepsilon_H)|u_iv_i|+O(\varepsilon_H^2)+O(\varepsilon_F). 
\end{aligned}
\label{equation: production error}
\end{equation}

Notice that the rounding function does not change the sign of the number. Therefore, relative error of the product is:
\begin{equation}
\begin{aligned}
    \varepsilon(u_iv_i) &= \dfrac{|u_iv_i-R_F(R_H(u_i)R_H(v_i))|}{|u_iv_i|} \\
    &=\dfrac{||u_iv_i|-|R_F(R_H(u_i)R_H(v_i))||}{|u_iv_i|} \\
    &\leq 2\varepsilon_H+O(\varepsilon_H^2)+O(\varepsilon_F). \label{equation: relative error of product}
\end{aligned}
\end{equation}

To simply the proof, we represent $a_i$ as the accumulator on Tensor Cores:
\begin{equation}
    a_i = R_F(R_H(u_i)R_H(v_i))=u_iv_i(1 \pm \varepsilon(u_iv_i)) \label{equation: accumulator}
\end{equation}

We assume that the addition is performed from left to right. Since the sum of each two values are stored in FP32 as a new accumulator, the computed cumulative sum $s_i$ of $\{a_i\}_{i=1}^n$ is:

\begin{equation}
    s_1 = R_F(a_1)=a_1,\quad s_i = R_F(s_{i-1}+a_i).
\end{equation}

Therefore, the computed value of $\beta$ on Tensor Cores is:
\begin{equation}
    \hat\beta=R_F(R_F(\cdots R_F(R_F(a_1+a_2)+a_3)\cdots)+a_n)=s_n.
\end{equation}

Next, we will prove that $s_i = \sum_{j=1}^ia_j+O(\varepsilon_F)$ inductively:
\begin{enumerate}
    \item It is obvious that $s_1=a_1 = a_1+O(\varepsilon_F)$;
    \item For a positive integer $i<n$, we assume that $s_{i} = \sum_{j=1}^ia_j+O(\varepsilon_F)$. According to Proposition \ref{proposition 0},
    \begin{equation}
        s_{i+1}=R_F(s_i+a_i)=s_i+a_i+O(\varepsilon_F)=\sum_{j=1}^{i+1}a_j+O(\varepsilon_F).
    \end{equation}
\end{enumerate}
According to Equation (\ref{equation: accumulator}),
\begin{equation}
    \hat{\beta}=s_n=\sum_{i=1}^na_i+O(\varepsilon_H) =\sum_{i=1}^n(u_iv_i \pm \varepsilon(u_iv_i))+O(\varepsilon_F).
\end{equation}

To sum up, the absolute error of $\beta$ satisfies
\begin{equation}
    \begin{aligned}
        |\hat\beta-\beta| &=\left|\sum_{i=1}^nu_iv_i(1\pm \varepsilon(u_iv_i))+O(\varepsilon_F)-\sum_{i=1}^nu_iv_i\right| \\
        &=\left|\sum_{i=1}^n\pm u_iv_i\varepsilon(u_iv_i)+O(\varepsilon_F)\right| \\
        &=\left|\sum_{i=1}^n\pm u_iv_i(2\varepsilon_H+O(\varepsilon_H^2)+O(\varepsilon_F))+O(\varepsilon_F)\right| \\
        &=\left|\sum_{i=1}^n\pm 2u_iv_i\varepsilon_H+O(\varepsilon_H^2)+O(\varepsilon_F)\right| \\
        &\leq2\varepsilon_H\sum_{i=1}^n|u_iv_i|+O(\varepsilon_H^2+\varepsilon_F).
    \end{aligned}
\end{equation}

The proof of Proposition \ref{proposition: dot product} is completed here.

\subsection{Absolute Error of $\beta$ on Global Coordinates}
\label{subsec: global err}

This section discusses the absolute of exponent $\beta$ on global coordinates.

The absolute error of $\beta$ is dependent on $\mathbf{v}$, and $\mathbf{v}$ is depend on the parameters of projected Gaussian $G'$. Therefore, we discusses the range of Gaussians' parameters at first.

We represent the projected Gaussian as $G'=(\mu',\Sigma',o,c)$ with $\mu'=(\mu'_x,\mu'_y)$ and symmetric $\Sigma'^{-1}=\begin{bmatrix}\sigma_{11}&\sigma_{12}\\\sigma_{21}&\sigma_{22}\end{bmatrix}$ satisfying $\sigma_{12}=\sigma_{21}$. The alpha value at pixel $p=(p_x,p_y)$ is $\alpha=\mathrm{e}^\beta=\mathrm{e^{\mathbf{u^T}\mathbf{v}}}$ as Section 4.2 discussed. 

The width and height of rendered image are represented as $w$ and $h$ respectively.

\paragraph{Means $\mu'$:} To simplify the analysis, we only consider Gaussians whose center within the screen space. Therefore, $\mu'\in[0,w]\times[0,h]$.

\paragraph{Opacity $o$:} The fragment with alpha value below $\frac{1}{255}$ will be culled. Since $\alpha\leq o$ since $\Sigma'^{-1}$ is semi-definite, $o \in [\frac{1}{255},1]$.

\paragraph{Inverse Covariance $\Sigma'^{-1}$:} The covariance $\Sigma'$ identifies the shape of the ellipse on the screen space. If $\lambda_1$ and $\lambda_2$ are the two eigenvalues of $\Sigma'$, then length of two semi-axes of the ellipse are $\lambda_1$ and $\lambda_2$. 

We assume the length of semi-axis is larger than 0.5, otherwise the Gaussian will not be rendered:
\begin{equation}
    \lambda_1,\lambda_2\geq0.5.
\end{equation}
Then the eigenvalues of the inverse covariance $\Sigma'^{-1}$ satisfies:
\begin{equation}
    0<\lambda_1^{-1}+\lambda_2^{-1}\leq4,\quad0<\lambda_1^{-1}\lambda_2^{-1}\leq 4.
\end{equation}

Consider the characteristic polynomial of $\Sigma^{\prime-1}$:
\begin{equation}
    f(\lambda^{-1})=|\lambda^{-1} I-\Sigma^{-1}|=\lambda^{-2}-(\sigma_{11}+\sigma_{22})\lambda^{-1}+\sigma_{11}\sigma_{22}-\sigma_{12}^2. \label{equation: polynomial}
\end{equation}

These two eigenvalues satisfy $f(\lambda^{-1})=0$. Applying the Vieta's Formulas,
\begin{equation}
    0<\sigma_{11}+\sigma_{22}\leq 4,\quad0<\sigma_{11}\sigma_{22}-\sigma_{12}^2\leq 4.
\end{equation}

Then,
\begin{equation}
\sigma_{12}^2<\sigma_{11}\sigma_{22}\leq\left(\frac{\sigma_{11}+\sigma_{22}}{2}\right)^2\leq4.
\end{equation}
Therefore,
\begin{equation}
0<\sigma_{11}<4,\quad -2\leq\sigma_{12}\leq 2,\quad0<\sigma_{22}<4.
\end{equation}

According to Proposition \ref{proposition: dot product}, the absolute error of $\beta$ on \textit{Frag2Mat} is:
\begin{equation}
    |\hat\beta-\beta|\leq 2\varepsilon_H\sum_{i=0}^5|u_iv_i|+O(\varepsilon_H^2+\varepsilon_F).
\end{equation}

Applying the global coordinate, the coordinate of pixel satisfies $p=(p_x,p_y)\in[0,w]\times[0,h]$.

Then,
\begin{gather}
|u_0v_0|=\left|\ln(o)-\frac{1}{2}\sigma_{11}\mu_x^2-\sigma_{12}\mu_x\mu_y-\sigma_{22}\mu_{y}^2\right|=O(w^2+wh+h^2), \\
|u_1v_1|=|p_x(\sigma_{11}\mu_x+\sigma_{12}\mu_y)|=O(w^2+wh), \\
|u_2v_2|=|p_y(\sigma_{12}\mu_x+\sigma_{22}\mu_y)|=O(wh+h^2), \\
|u_3v_3|=\left|p_x^2(-\frac{1}{2}\sigma_{11})\right|=O(w^2),\\
|u_4v_4|=|p_xp_y(-\sigma_{12})|=O(wh), \\
|u_5v_5|=\left|p_y^2(-\frac{1}{2}\sigma_{22})\right|=O(h^2).
\end{gather}

Therefore, the absolute error of $\beta$ satisfies quadratic
\begin{equation}
    |\hat\beta-\beta|=O(w^2+wh+h^2)\varepsilon_H+O(\varepsilon_H^2+\varepsilon_F).
\end{equation}

\subsection{Absolute Error of $\beta$ on Local Coordinates}
This section discusses the absolute of exponent $\beta$ on local coordinates. 

At first, we discuss the range of Gaussians' parameters as Section \ref{subsec: global err}. In this section, $p(\mathcal{T})$ is the center of tile $\mathcal{T}$, and the size of each tile is $16\times 16$.

\paragraph{Means $\Delta\mu'$:} Since $\mu',p(\mathcal{T})\in[0,w]\times[0,h]$, $\Delta\mu'=\mu'-p(\mathcal{T})\in[-w,w]\times[-h,h]$.

\paragraph{Opacity $o$:} The opacity is the same as Gaussians on global coordinates. $o \in [\frac{1}{255},1]$.

\paragraph{Inverse Covariance $\Sigma'^{-1}$:}. The covariance is the same as Gaussians on global coordinates: $0<\sigma_{11},\sigma_{12}<4,-2\leq\sigma_{12}\leq2$.

All pixels on local coordinates is within a tile, then $\Delta p=(\Delta p_x,\Delta p_y)\in[-8,8]^2$.

Therefore,
\begin{gather}
|u_1v_1|=|\Delta p_x(\sigma_{11}\Delta\mu_x+\sigma_{12}\Delta\mu_y)|=O(w+h), \\
|u_2v_2|=|\Delta p_y(\sigma_{12}\Delta\mu_x+\sigma_{22}\Delta\mu_y)|=O(w+h), \\
|u_3v_3|=\left|\Delta p_x^2(-\frac{1}{2}\sigma_{11})\right|=O(1),\\
|u_4v_4|=| \Delta p_x \Delta p_y(-\sigma_{12})|=O(1), \\
|u_5v_5|=\left| \Delta p_y^2(-\frac{1}{2}\sigma_{22})\right|=O(1).
\end{gather}

According to Proposition \ref{proposition: dot product},
\begin{equation}
    |\hat\beta-\beta|\leq(2|u_0v_0|+O(w+h))\varepsilon_H+O(\varepsilon_H^2+\varepsilon_F).
\end{equation}

We only care the precision of fragments which are not culled, then $-\ln(255)\leq\beta<0$. Since $\beta=\sum_{i=0}^5u_iv_i$, then $u_0v_0=\beta-\sum_{i=1}^5u_iv_i$. Therefore,
\begin{equation}
    |u_0v_0|\leq|\beta|+\sum_{i=1}^5|u_iv_i|=O(w+h).
\end{equation}

In conclusion, the absolute error of $\beta$ satisfies
\begin{equation}
    |\hat\beta-\beta|\leq O(w+h)\varepsilon_H+O(\varepsilon_H^2+\varepsilon_F).
\end{equation}

\section{Notations}
To clearly present the various parameters and their definitions involved in this paper, the following parameter table lists each symbol along with its corresponding meaning, facilitating readers' understanding of the subsequent analysis and computational processes.
\begin{table*}[htbp]
    \centering
    \begin{tabular}{cl|cl}
    \hline
    Notation & Meaning & Notation & Meaning \\
    \hline
    $\mathcal{G}$ & Set of all 3D Gaussian & $G$ & 3D Gaussian\\
    $P$ & Number of 3D Gaussians &  $\mu$ & Mean of 3D Gaussian \\
    $\Sigma$ & Covariance of Covariance & $o$ & Opacity of 3D Gaussian \\
    $c$ & Color of 3D Gaussian & $W$ & Matrix projecting 3D position into 2D plane \\
    $J$ & Jacobian of projection & $\mu^\prime=(\mu^\prime_x,\mu^\prime_y)$ & Mean of projected Gaussian \\
    $\Sigma^\prime = \begin{bmatrix}\sigma_{11}&\sigma_{12}\\\sigma_{21}&\sigma_{22}\end{bmatrix}^{-1}$ & Covariance of projected Gaussian & $G^\prime$ & Projected Gaussian \\
    $\mathcal{G}^\prime$ & Set of projected Gaussians & $p=(p_x,p_y)$ & Pixel \\
    $\mathcal{T}$ & Tile, containing pixels & $\mathcal{S}$ & Screen space, containing all tiles \\
    $C$ & Rendered pixel color & $\alpha$ & Alpha value of fragments. \\
    $T$ & Transmissivity of fragments. & $n(\mathcal{T})$& Number of splats on tile $\mathcal{T}$\\
    $N$ & Number of splats. & $\mathcal{E}$ & Complexity of the pipeline in 3DGS \\
    $w$ & Width of the screen & $h$ & Height of the screen \\
    $\mathcal{E}_p$ & Complexity of preprocessing & $\mathcal{E}_s$ & Complexity of sorting \\
    $\mathcal{E}_a$ & Complexity of alpha-blending & $k_\alpha$ & Computation amount in alpha computation \\
    $k_{\text{cull}}$ & Computation amount in culling & $k_{\text{blend}}$ & Computation amount in blending   \\
    $f_\text{cull}$ & Number of culled fragments& $f_{\text{blend}}$ & Number of blended (shaded) fragments. \\
    $f_\text{skip}$ & Number of skipped fragments & $\mathcal{C}$ & Total computation amount in alpha-blending. \\
    $\mathbf{u}$ & Pixel vector & $\mathbf{v}$  &Gaussian vector \\
    $(p_x,p_y)$ & Coordinates of pixel $p$ & $\beta$ & $\ln(\alpha)$, the exponent \\
    $\mathbf{U}$ & Pixel matrix & $\mathbf{V}$ & Gaussian matrix \\
    $\mathbf{B}$ & Exponent matrix & $p(\mathcal{T})$ & Center of tile $\mathcal{T}$ \\
    $\Delta p=(\Delta p_x,\Delta p_y).$ & Pixel on local coordinates & $\Delta\mu^\prime=(\Delta\mu_x',\Delta\mu_y')$ & Projected mean on local coordinates \\
    $\varepsilon_H$ & Machine epsilon of FP16 or TF32 & $\varepsilon_F$ & Machine epsilon of FP32 \\
    $R_H$ & Rounding function of FP16 or TF32 & $R_F$ & Rounding function of FP32 \\
     $\hat{\beta}$ & Computed value of $\beta$ with Tensor Cores & $\lambda_1,\lambda_2$ & Two eigenvalues of $\Sigma^\prime$ \\
    \end{tabular}
    \caption{Notations in this paper.}
    \label{tab:my_label}
\end{table*}

\end{document}